\documentclass[12pt]{article}
\usepackage{amssymb,amsmath,graphicx,color}
\newcommand{\be}{\begin{equation}}
\newcommand{\ee}{\end{equation}}
\newcommand{\bear}{\begin{eqnarray}}
\newcommand{\eear}{\end{eqnarray}}
\newcommand{\ba}{\begin{array}}
\newcommand{\ea}{\end{array}}
\newcommand{\nn}{\nonumber}

\begin{document}
\vspace{9mm}

\begin{center}
{{{\Large \bf On the Vacua of Mass-deformed Gaiotto-Tomasiello
Theories}
}\\[17mm]
O-Kab Kwon$^{1}$,~~D.~D. Tolla$^{1,2}$\\[3mm]
{\it $^{1}$Department of Physics,~BK21 Physics Research Division,
~Institute of Basic Science,\\
$^{2}$University College,\\
Sungkyunkwan University, Suwon 440-746, Korea}\\[2mm]
{\tt okab@skku.edu,~ddtolla@skku.edu} }
\end{center}

\vspace{20mm}

\begin{abstract}
We write explicit Lagrangian and supersymmetry transformation rules
using the component fields in the ${\cal N}=2,3$ GT theories. In the
component field expansion, the manifestation of an additional ${\cal
N}=1$ supersymmetry is verified in the ${\cal N}=3$ GT theory. We
find maximal supersymmetry preserving mass-deformation of the GT
theories and their classical supersymmetric discrete vacua. Some
interesting aspects of the set of discrete vacua are discussed in
comparison with the ABJM case.

\end{abstract}

\newpage

\tableofcontents

\section{Introduction}

Massive type IIA supergravity~\cite{Romans:1985tz} has many
supersymmetric and nonsupersymmetric solutions of the form
AdS$_4\times {\cal M}_6$, where ${\cal M}_6$ is a six dimensional
manifold. Some of the nonsupersymmetric solutions were already found
in \cite{Romans:1985tz}, while the supersymmetric ones were not
known until recently. The first such solution was ${\cal N}=1$
solution constructed  in \cite{Behrndt:2004mj} and later generalized
in \cite{Lust:2004ig,Tomasiello:2007eq,Koerber:2008rx}. Based on these works,
${\cal N}=2$ solutions were found by compactifying
${\rm AdS}_4\times  M^{(1,1,1)}$ solutions and introducing mass
deformation~\cite{Petrini:2009ur}
(see \cite{Lust:2009mb,Aharony:2010af} for more general consideration).
Other family of
${\cal N}=2$ solutions, including massive deformation of the
compactified ${\rm AdS}_4\times Q^{(1,1,1)}$ solution, were also
constructed~\cite{Tomasiello:2010zz}.

In relation with the dual three dimensional superconformal field
theories for the above solutions, Gaiotto and
Tomasiello~\cite{Gaiotto:2009mv} considered some deformations in the
Aharony-Bergman-Jafferis-Maldacena (ABJM)
theory~\cite{Aharony:2008ug} such that the sum of Chern-Simons (CS)
levels for the two gauge fields is not zero.\footnote{The shift in Chern-Simons levels in the presence of D8-branes was also discussed in \cite{Fujita:2009kw}.} The deformations yield in different superconformal CS-matter theories with ${\cal
N}=0,1,2,$ and $3$ supersymmetries and SO(6), SO(5), SO(2)$_{\rm
R}\times {\rm SO}(4)$, and SO(3)$_{\rm R}\times {\rm SO}(3)$ global
symmetries, respectively. We refer to  these theories as GT theories
in the sequel. The authors argued that the deformed theories are
dual to massive type  IIA supergravity with the Roman mass
parameter, which is understood as the RR zero-form flux $F_0$, is
identified as the sum of CS levels of the two gauge fields $(F_0 =
k_1 + k_2)$~\cite{Gaiotto:2009mv,Gaiotto:2009yz}. See also \cite{Fujita:2009kw}. In particular, the gravity dual for the ${\cal N}=0$ GT theory was identified as
AdS$_4\times\mathbb{CP}^3$ solution in \cite{Romans:1985tz} and the
solutions in \cite{Behrndt:2004mj,Tomasiello:2007eq,Koerber:2008rx}
are conjectured to be dual to ${\cal N}=1$ GT theory. As a further
evidence, a brane configuration for the ${\cal N} = 0$ theory was
proposed in Type IIB string theory~\cite{Bergman:2010xd}, by
introducing D7 branes to the brane configuration of the ABJM
theory~\cite{Aharony:2008ug}. Here $F_0$ is identified with the
number of D7-branes. The gravity duals for the ${\cal N}=2$ and
${\cal N}=3$ GT theories with small $F_0$ were obtained in
~\cite{Gaiotto:2009yz}
  by using first order perturbation for the AdS$_4\times\mathbb{CP}^3$ solution,
which is dual to the ${\cal N}=6$ ABJM theory. Some nonperturbative
aspects of ${\cal N}=2,3$ GT theories were  also discussed by
calculating partition
functions~\cite{Suyama:2010hr,Jafferis:2011zi,Suyama:2011yz}.

An important missing point in this subject is that we have no known
M-theory interpretation and dual eleven dimensional gravity for the
GT theories. The reason is that the existence of the M-theory limit
of the massive type IIA string theory is unclear. Recently, there
was a conjecture of nonexistence of that
limit~\cite{Aharony:2010af}, though the authors only considered
weakly-curved solutions in massive type IIA gravity. The validity of
this conjecture in the strongly-curved theory is unclear  and needs
further investigation.

On the other hand maximal supersymmetry preserving nonconformal
deformations of the ABJM theory were found in
\cite{Hosomichi:2008jb,Gomis:2008vc} and  interpreted as the
worldvolume theory of multiple M2-branes with a constant transverse
four-form field strength~\cite{Lambert:2009qw,Kim:2010hj}. An
important feature of this theory is that it has many classical
discrete supersymmetric vacua~\cite{Gomis:2008vc} with their number
much larger than the number expected from the dual gravity solution
which is the Lin-Lunin-Maldacena (LLM)
geometry~\cite{Lin:2004nb}\footnote{See also \cite{Bena:2004jw}.}
for the CS level $k=1$. The latter problem was recently resolved by
realizing that many of the classical supersymmetric vacua
dynamically breaks supersymmetry and results in the expected
number~\cite{Kim:2010mr}. This gauge/gravity duality relation
between the mass-deformed ABJM theory and the LLM geometry was
extended to generic $k$ in \cite{Cheon:2011gv}  and the role of $k$
was identified as the $\mathbb{Z}_k$ quotients of the LLM geometry.

In this paper, we consider the  mass deformation of the ${\cal
N}=2,3$ GT theories. We verify that like the ABJM theory the GT
theories allow maximal supersymmetry preserving mass deformations.
We solve the vacuum equation and find discrete supersymmetric vacua
which are similar in structure and property to the discrete
solutions of the mass-deformed ABJM
theory~\cite{Gomis:2008vc,Kim:2010mr,Cheon:2011gv}. More precisely,
the solutions are represented in terms of $n\times(n+1)$ or
$(n+1)\times n$ irreducible blocks. One basic difference is that
there are overall coefficients which depend on the ratio of the CS
levels $t=-\frac{k_2}{k_1}$ and the size of the irreducible block
$n$. For some special values of $t$, the coefficients of some blocks
are singular and those blocks are not allowed. This fact reduces the
total number of vacua as compared to the case of ABJM theory.

This paper is organized as follows. In section 2 we write the ${\cal
N}=2,3$ GT Lagrangian in terms of component fields and the
corresponding supersymmetry transformation rules. In section 3 we
obtain the maximal supersymmetry preserving mass deformation of
these theories. In section 4 we solve the vacuum equations for the
mass-deformed theories and find sets of discrete vacua. Section 5
includes conclusion and feature directions.

\section{${\cal N}=2$ and ${\cal N}=3$  GT Theories}\label{sectionN=23}

The actions for these theories were written in \cite{Gaiotto:2009mv}
using the ${\cal N}=2$ superfield formulation.
In this section, we will give the Lagrangians in terms of component fields
and find the corresponding supersymmetric transformation rules.

For convenience we review the ${\cal N}=2$ superfield formulation of
the GT theories with gauge group U($N)_{k_1}$$\times$U($N)_{k_2}$.
The action is given by\footnote{We mainly follow the notations of
\cite{Benna:2008zy}.}
\begin{align}\label{sfN=2GT}
S_{{\cal N}=2} = -\frac{ik_1}{8\pi} S_{{\rm CS}}({\cal V}_1)
-\frac{ik_2}{8\pi} S_{{\rm CS}}({\cal V}_2) + S_{{\rm mat}} +
S_{{\rm pot}},
\end{align}
where
\begin{align}\label{sfN=2GT2}
&S_{{\rm CS}}({\cal V}) = \int d^3 x d^4\theta\int_0^1 dt\, {\rm
tr}\big[ {\cal V}\bar D^\alpha\big(e^{t{\cal V}} D_\alpha e^{-t{\cal
V}}\big)\big],
\nonumber \\
&S_{{\rm mat}}= \int d^3x d^4\theta\,{\rm tr}\big[-\bar {\cal Z}_A
e^{-{\cal V}_1} {\cal Z}^A e^{{\cal V}_2} - \bar{\cal W}^A e^{-{\cal
V}_2} {\cal W}_A e^{{\cal V}_1}\big],
\nonumber \\
&S_{{\rm pot}}= c_1\int d^3 x d^2\theta\, {\rm tr}\big[ {\cal
Z}^A{\cal W}_A{\cal Z}^B{\cal W}_B\big] + c_1\int d^3x d^2\bar\theta
\,{\rm tr}\big[ \bar{\cal W}^A\bar{\cal Z}_A\bar{\cal W}^B\bar{\cal
Z}_B\big]
\nonumber \\
&~~~~~~+c_2\int d^3 x d^2\theta\, {\rm tr}\big[ {\cal W}_A{\cal
Z}^A{\cal W}_B{\cal Z}^B\big] + c_2\int d^3x d^2\bar\theta \,{\rm
tr}\big[ \bar{\cal Z}_A\bar{\cal W}^A\bar{\cal Z}_B\bar{\cal W}^B\big]
\end{align}
with $A,B=1,2$ and arbitrary real numbers $c_1$, $c_2$. The component field
expansions for the chiral superfields, ${\cal Z}^A$, ${\cal W}_A$,
and anti-chiral ones, $\bar{\cal Z}_A$, $\bar{\cal W}^A$, are
given by
\begin{align}
{\cal Z}^A &= Z^A(y) + \sqrt{2}\theta\xi^A (y) + \theta^2 F^A(y),
\quad \bar{\cal Z}_A = Z_A^\dagger(y^\dagger) -
\sqrt{2}\bar\theta\xi_A^\dagger (y^\dagger) - \bar\theta^2
F_A^\dagger(y^\dagger),
\nonumber \\
{\cal W}_A &= W_A(y) + \sqrt{2}\theta\omega_A (y) + \theta^2 G_A(y),
\quad \bar{\cal W}^A = W^{\dagger A}(y^\dagger) -
\sqrt{2}\bar\theta\omega^{\dagger A} (y^\dagger) - \bar\theta^2
G^{\dagger A}(y^\dagger),
\end{align}
where the supercoordinate $y$ is defined as
\begin{align}
y^\mu = x^\mu -i\theta\gamma^\mu\bar\theta,\qquad  y^{\dagger \mu} =
x^\mu +i\theta\gamma^\mu\bar\theta.
\end{align}
${\cal Z}^A$ and $\bar{\cal W}^A$ are in the bifundamental
representations, while $\bar{\cal Z}_A$ and ${\cal W}_A$ are in the
anti-bifundamental representations of the gauge group. The component
field expansions of the vector superfields ${\cal V}_1$ and ${\cal
V}_2$ in Wess-Zumino gauge are
\begin{align}
{\cal V}_1&= 2i\theta\bar\theta\sigma_1 (x) -
2\theta\gamma^\mu\bar\theta A_\mu(x) +
\sqrt{2}i\theta^2\bar\theta\bar\chi_1 (x) - \sqrt{2}
i\bar\theta^2\theta\chi_1 + \theta^2\bar\theta^2 D_1(x),
\nonumber \\
{\cal V}_2&= 2i\theta\bar\theta\sigma_2 (x) -
2\theta\gamma^\mu\bar\theta \hat A_\mu(x) +
\sqrt{2}i\theta^2\bar\theta\bar\chi_2 (x) - \sqrt{2}
i\bar\theta^2\theta\chi_2 + \theta^2\bar\theta^2 D_2(x).
\end{align}
Some conventions of the ${\cal N}=2$ superspace are given in appendix \ref{CV-UF}.

For generic $c_i$, the theory \eqref{sfN=2GT} has ${\cal N}=2$ supersymmetry
(SO(2) R-symmetry) and SU(2) flavor symmetry. For a particular case of $c_1=-c_2=c$,
the superpotential in \eqref{sfN=2GT2} can be rewritten as
\begin{align}
S_{{\rm pot}}&= c\int d^3 x d^2\theta\,\epsilon_{AB}\epsilon^{CD}
{\rm tr}\big[ {\cal Z}^A{\cal W}_C{\cal Z}^B{\cal W}_D\big] + c\int d^3x d^2\bar\theta
\,\epsilon^{AB}\epsilon_{CD}{\rm tr}\big[
\bar{\cal W}^C\bar{\cal Z}_B\bar{\cal W}^D\bar{\cal Z}_B\big].
\end{align}
The supersymmetry of this theory remains ${\cal N}=2$, however, the
flavor symmetry is enhanced to SU(2)$\times$SU(2). On the other
hand, if we choose $c_i = \frac{2\pi}{k_i}$, the supersymmetry is
enhanced to ${\cal N}=3$, while the flavor symmetry remains
SU(2)~\cite{Gaiotto:2009mv,Gaiotto:2009yz}. In addition, if $F_0=0$,
the supersymmetry is enhanced to ${\cal
N}=6$~\cite{Aharony:2008ug,Aharony:2008gk} and to ${\cal N}=8$ for
$k=1,2$~\cite{Gustavsson:2009pm,Bashkirov:2010kz}.

\subsection{${\cal N}=2$}\label{2.1}

In component field notation the ${\cal N}=2$ Lagrangian can be written as
\begin{align}\label{N=2GT}
{\cal L}_{{\cal N}=2}={\cal L}_{0}+{\cal L}_{\rm CS}+{\cal L}^D_{\rm
ferm}+{\cal L}^F_{\rm ferm}+{\cal L}^D_{\rm bos}+{\cal L}^F_{\rm
bos}
\end{align}
where
\begin{align}\label{LLCS}
{\cal L}_{0}&={\rm tr}\big[-D_\mu Z^\dagger_AD^\mu Z^A-D_\mu
W^{\dagger A}D^\mu W_A+i\xi^{\dagger}_A\gamma^\mu
D_\mu\xi^A+i\omega^{\dagger A}\gamma^\mu D_\mu\omega_A\big],
\nonumber \\
{\cal L}_{\rm CS}&=\frac{k_1}{4\pi}\epsilon^{\mu\nu\rho} {\rm
tr}\Big(A_\mu\partial_\nu A_\rho+\frac{2i}3A_\mu A_\nu
A_\rho\Big)+\frac{k_2}{4\pi}\epsilon^{\mu\nu\rho} {\rm tr} \Big(\hat
A_\mu\partial_\nu \hat A_\rho +\frac{2i}3\hat A_\mu \hat A_\nu \hat
A_\rho\Big),
\end{align}
\begin{align}\label{LfermDF}
{\cal L}^D_{\rm ferm}&=-\frac{2\pi i}{k_1}{\rm tr}\Big[
(\xi^A\xi^\dagger_A-\omega^{\dagger
A}\omega_A)(Z^BZ^\dagger_B-W^{\dagger B}W_B) +2 (Z^A\xi^\dagger_A
-\omega^{\dagger A}W_A)(\xi^BZ^\dagger_B-W^{\dagger B}\omega_B)\Big]
\nonumber \\
&~~~-\frac{2\pi i}{k_2}{\rm
tr}\Big[(\xi^\dagger_A\xi^A-\omega_A\omega^{\dagger
A})(Z^{\dagger}_BZ^B-W_BW^{\dagger B})
+2(Z^\dagger_A\xi^A-\omega_AW^{\dagger A})(\xi^{\dagger}_BZ^B
-W_B\omega^{\dagger B})\Big],
\nonumber \\
{\cal L}^F_{\rm ferm} &= -c_1\,{\rm tr}\big(Z^A\omega_AZ^B\omega_B
+\xi^AW_A\xi^BW_B +2Z^AW_A\xi^B\omega_B +2Z^A\omega_A\xi^BW_B
\nonumber \\
&\hskip 1.5cm - \omega^{\dagger A}Z_A^\dagger\omega^{\dagger
B}Z_B^\dagger - W^{\dagger A}\xi_A^\dagger W^{\dagger
B}\xi_B^\dagger -2\omega^{\dagger A}\xi_A^\dagger W^{\dagger
B}Z_B^\dagger - 2 W^{\dagger A}\xi_A^\dagger\omega^{\dagger
B}Z_B^\dagger\big)
\nonumber \\
&~~~-c_2\,{\rm tr}\big(\omega_AZ^A\omega_BZ^B+W_A\xi^AW_B\xi^B
+2\omega_AZ^AW_B\xi^B +2W_AZ^A\omega_B\xi^B
\nonumber \\
&\hskip 1.5cm - Z_A^\dagger\omega^{\dagger
A}Z_B^\dagger\omega^{\dagger B} -\xi_A^\dagger W^{\dagger
A}\xi_B^\dagger W^{\dagger B} -2\xi_A^\dagger W^{\dagger A}
Z_B^\dagger\omega^{\dagger B} -2\xi_A^\dagger\omega^{\dagger A}
Z_B^\dagger W^{\dagger B}  \big),
\end{align}
and
\begin{align}\label{LbosDF}
{\cal L}_{{\rm bos}}^D &= -\frac{4\pi^2}{k_1^2}{\rm
tr}\Big[\big(Z^AZ_A^\dagger + W^{\dagger
A}W_A\big)\big(Z^BZ_B^\dagger - W^{\dagger B}W_B\big)
(Z^CZ_C^\dagger - W^{\dagger C}W_C\big)\Big]
\nonumber \\
&~~~-\frac{8\pi^2}{k_1k_2}{\rm tr}\Big[\big(Z^AZ_A^\dagger -
W^{\dagger A}W_A\big)Z^B (Z_C^\dagger Z^C - W_CW^{\dagger
C}\big)Z_B^\dagger
\nonumber \\
&~~~~~~~~~~~~~~~+\big(Z^AZ_A^\dagger - W^{\dagger A}W_A\big)W^{\dagger B}
(Z_C^\dagger Z^C - W_CW^{\dagger C}\big)W_B\Big]
\nonumber \\
&~~~-\frac{4\pi^2}{k_2^2}{\rm tr}\Big[\big(Z_A^\dagger Z^A +
W_AW^{\dagger A}\big)\big(Z_B^\dagger Z^B - W_BW^{\dagger B}\big)
(Z_C^\dagger Z^C - W_CW^{\dagger C}\big)\Big],
\nonumber\\
{\cal L}_{{\rm bos}}^F &= -4{\rm tr}\Big[\big(c_1W_AZ^BW_B + c_2 W_B
Z^B W_A\big) \big(c_1W^{\dagger C}Z_C^\dagger W^{\dagger A} + c_2
W^{\dagger A}Z_C^\dagger W^{\dagger C}\big)
\nonumber \\
&~~~~~~~~~~~+\big(c_1Z^B W_BZ^A + c_2 Z^A W_B Z^B\big)\big(c_1 Z_A^\dagger
W^{\dagger C}Z_C^\dagger + c_2 Z_C^\dagger W^{\dagger C}
Z_A^\dagger\big)\Big].
\end{align}

In generic case of $c_i$, the ${\cal N}=2$ supersymmetric
transformation rules for the component fields are given by
\begin{align}\label{N=2susy}
&\delta_{\epsilon} Z^A = i\bar\epsilon\xi^A,\quad \delta_{\epsilon}
Z_A^\dagger = i\xi_A^\dagger\epsilon,
\nonumber \\
&\delta_{\epsilon} W_A = i\bar\epsilon \omega_A,\quad
\delta_{\epsilon} W^{\dagger A} = i\omega^{\dagger A}\epsilon,
\nonumber \\
&\delta_{\epsilon}\xi^A = -D_\mu Z^A \gamma^\mu\epsilon
-\sigma_1Z^A\epsilon + Z^A\sigma_2\epsilon
-2i\bar\epsilon\big(c_1W^{\dagger B}Z_B^\dagger W^{\dagger A} + c_2
W^{\dagger A} Z_B^\dagger W^{\dagger B}\big),
\nonumber \\
&\delta_{\epsilon} \xi_A^\dagger = \bar\epsilon\gamma^\mu D_\mu
Z_A^\dagger -\bar\epsilon Z_A^\dagger\sigma_1 +\bar\epsilon\sigma_2
Z_A^\dagger + 2i\big(c_1W_AZ^BW_B + c_2 W_B Z^B W_A\big)\epsilon,
\nonumber\\
&\delta_{\epsilon}\omega_A = - D_\mu W_A\gamma^\mu\epsilon
+W_A\sigma_1\epsilon -\sigma_2 W_A\epsilon
-2i\bar\epsilon\big(c_1Z_A^\dagger W^{\dagger B} Z_B^\dagger +
c_2Z_B^\dagger W^{\dagger B} Z_A^\dagger \big),
\nonumber \\
&\delta_{\epsilon}\omega^{\dagger A} = \bar\epsilon\gamma^\mu D_\mu
W^{\dagger A} +\bar\epsilon\sigma_1 W^{\dagger A} - \bar\epsilon
W^{\dagger A}\sigma_2 + 2i\big( c_1 Z^B W_B Z^A + c_2 Z^A W_B
Z^B\big)\epsilon,
\nonumber \\
&\delta_{\epsilon} A_\mu =
\frac12\big(\bar\epsilon\gamma_\mu\bar\chi_1 +
\chi_1\gamma_\mu\epsilon\big), \quad \delta_{\epsilon}\hat A_\mu
=\frac12\big(\bar\epsilon\gamma_\mu\bar\chi_2 +
\chi_2\gamma_\mu\epsilon\big),
\end{align}
where the supersymmetry parameters $\epsilon$ and $\bar\epsilon$ are complex two
component spinor and its complex conjugate, respectively. Here we also defined
\begin{align}\label{auxilflds}
&\sigma_1 \equiv \frac{2\pi}{k_1}\big(Z^BZ_B^\dagger - W^{\dagger
B}W_B\big), \qquad \sigma_2 \equiv -\frac{2\pi}{k_2}\big(Z_B^\dagger
Z^B - W_BW^{\dagger B}\big),
\nonumber \\
&\chi_1 \equiv -\frac{4\pi}{k_1}\big(Z^A\xi_A^\dagger-\omega^{\dagger A}W_A\big),
\qquad \chi_2 \equiv \frac{4\pi}{k_2} \big(\xi_A^\dagger Z^A - W_A\omega^{\dagger A}\big).
\end{align}

The ${\cal N}=2$ supersymmetric parameter $\epsilon$ is inherited
from the original ${\cal N}=6$ supersymmetry in the ABJM theory. The
${\cal N}=6$ supersymmetric parameters $\omega^{AB}$, ($A=1,2,3,4$),
can be grouped as the parameters of ${\cal N}=2$ and those of ${\cal
N}=4$. The parameters of ${\cal N}=2$ are $\omega^{ab}$ $(a,b=1,2)$
and they are related to $\omega_{\dot a\dot b}$ ($\dot a,\dot
b=3,4$) by reality condition, while those of ${\cal N}=4$ are
$\omega^{a\dot b}$. The parameter $\epsilon$ in \eqref{N=2susy} is
identified with $\omega^{12}=\omega_{34}$. Therefore, the ${\cal
N}=4$ part in the ABJM theory are broken by introducing nonvanishing
$F_0$ with generic $c_i$ in the $F$-terms of the ${\cal N}=2$ GT
theory.

\subsection{${\cal N}=3$}
In this subsection, we find the additional ${\cal N}=1$
supersymmetry in the action \eqref{N=2GT} when $c_i =
\frac{2\pi}{k_i}$. This additional supersymmetry is slightly
different from the ${\cal N}=2$ of the previous section. For this
reason we will briefly summarize the invariance of the action under
this supersymmetry.

We start by noting that under this supersymmetry we have
\begin{align}
\delta_\eta {\cal L}_0+\delta_\eta^A {\cal L}_{\rm CS}=0,
\end{align}
where the supersymmetric variations are
\begin{align}\label{susyN=3-1}
\begin{array}{ll}\delta_\eta Z^A = -\eta\omega^{\dagger A},&
\delta_\eta Z_A^\dagger = -i\omega_A\eta,\\
\delta_\eta W_A = \eta\xi^\dagger_A,& \delta_\eta
W^{\dagger A} =i\xi^A\eta,\\
\delta_\eta\xi^A = iD_\mu W^{\dagger A} \gamma^\mu\eta,
& \delta_\eta \xi_A^\dagger = \eta\gamma^\mu D_\mu W_A \\
\delta_\eta\omega_A = -iD_\mu Z^\dagger_A\gamma^\mu\eta,
& \delta_\eta\omega^{\dagger A} = -\eta\gamma^\mu D_\mu Z^A,\\
\delta_\eta^A A_\mu = -\frac12\big(\eta\gamma_\mu\zeta_1 +
\bar\zeta_1\gamma_\mu\bar\eta\big), & \delta_\eta^A\hat A_\mu =
\frac12 \big(\eta\gamma_\mu\zeta_2 +
\bar\zeta_2\gamma_\mu\bar\eta\big).
\end{array}
\end{align}
Here we defined
\begin{align}
\zeta_1\equiv \frac{4\pi}{k_1}\big(\xi^AW_A + Z^A\omega_A\big),\qquad
\zeta_2\equiv\frac{4\pi}{k_2}\big(W_A\xi^A + \omega_AZ^A\big),
\end{align}
and the supersymmetric parameter $\eta$ is a two component complex
spinor. Later, this parameter will be constrained to give the ${\cal
N}=1$ parameter.

In order to complete the invariance of the action, we introduce an additional
transformation  for the fermions as
\begin{align}\label{susyN=3-2}
&\delta_\eta^{'}\xi^A = i\eta\sigma_1W^{\dagger A} - i\eta
W^{\dagger A}\sigma_2+\frac{4\pi i}{k_1}\eta W^{\dagger
B}Z_B^\dagger Z^A +\frac{4\pi i}{k_2}\eta Z^AZ_B^\dagger W^{\dagger
B},
\nonumber \\
&\delta_\eta^{'} \xi_A^\dagger =  -\eta W_A\sigma_1 +\eta\sigma_2
W_A-\frac{4\pi}{k_1}\eta Z^\dagger_A Z^BW_B-\frac{4\pi}{k_2} \eta
W_BZ^BZ^\dagger_A,
\nonumber \\
&\delta_\eta^{'}\omega_A = i\eta Z^\dagger_A\sigma_1 -i\eta\sigma_2
Z^\dagger_A-\frac{4\pi i}{k_1}\eta W_AW^{\dagger B}Z^\dagger_B
-\frac{4\pi i}{k_2}\eta Z^\dagger_BW^{\dagger B}W_A,
\nonumber \\
&\delta_\eta^{'}\omega^{\dagger A} = -\eta\sigma_1 Z^A +\eta
Z^A\sigma_2+\frac{4\pi}{k_1}\eta Z^BW_BW^{\dagger A}
 +\frac{4\pi}{k_2}\eta  W^{\dagger A}W_BZ^B.
\end{align}
Then we note that
\begin{align}\label{susyrel2}
\delta_\eta^A {\cal L}_0 + \delta_\eta^{'}{\cal L}_0 +\delta_\eta
{\cal L}_{{\rm ferm}}^{D} + \delta_\eta {\cal L}_{{\rm ferm}}^{F}
=0,
\end{align}
if the complex two component spinor $\eta$ satisfies the relation
\begin{align}
\eta=i\eta^\ast.
\end{align}
The remaining variations of the Lagrangian satisfy
\begin{align}\label{susyrel4}
\delta_\eta^{'} {\cal L}_{\rm ferm}^D+\delta_\eta^{'}{\cal L}_{\rm
ferm}^F+\delta_\eta {\cal L}_{\rm bos}^D+\delta_\eta {\cal L}_{\rm
bos}^F=0
\end{align}
without further constraint. This completes the verification of the
invariance of the action under the additional ${\cal N}=1$
supersymmetry.

We have relations similar to \eqref{susyrel2} and \eqref{susyrel4}
in the case of the ${\cal N}=2$ supersymmetric theory discussed in
the previous subsection. In that case,  the variations of the
$F$-term Lagrangians cancel with some part of variations of ${\cal
L}_0$ in \eqref{susyrel2} and each other in \eqref{susyrel4},
independent of the variations of the $D$-term Lagrangians. However,
in the case of the additional ${\cal N}=1$,  the supersymmetric
variations of the $D$-term and $F$-term Lagrangians are mixed. This
indicates that the additional ${\cal N}=1$ supersymmetry is
inherited from the ${\cal N}=4$ part of the original ABJM theory,
which mixes the component fields of the superfields ${\cal Z}^A$ and
${\cal W}_A$. For this reason, the supersymmetric parameter $\eta$
is inherited from the ${\cal N}=4$ supersymmetric parameter
$\omega^{a\dot b}$. In our conventions  $\eta=i\omega^{1\dot 2}$.

\section{Mass-deformed GT Theories}

It is well-known that the maximal supersymmetry preserving mass
deformation is possible in the ABJM
theory~\cite{Hosomichi:2008jb,Gomis:2008vc}. There are several
methods to obtain such mass-deformed theory, for instance,  ${\cal
N} = 1$ superfield formalism~\cite{Hosomichi:2008jb}, $D$-term and
$F$ -term deformations in ${\cal N} = 2$ superfield
formalism~\cite{Gomis:2008vc}. These different versions of
mass-deformed theories are actually equivalent since they are
connected by field redefinitions~\cite{Kim:2009ny}. Based on the
close relation between the ABJM and GT Lagrangians,  we expect that
such maximal supersymmetry preserving mass deformation can exist for
the supersymmetric GT theories as well. In this section,  we will
find such deformation for both ${\cal N}=2$ and ${\cal N}=3$ GT
theories.

\subsection{$D$-term deformation}
As in the ABJM theory the $D$-term deformation (FI deformation) is one way of obtaining
the supersymmetry preserving mass-deformation of the
GT theories.  In the ${\cal N}=2$ superfield formulation, the $D$-term deformation is given by
\begin{align}
S_D=-\frac{\mu}{4\pi}\int d^3xd^4\theta\,{\rm tr}\big(k_1{\cal
V}_1-k_2{\cal V}_2\big)=-\frac{\mu}{4\pi}\int d^3x\,{\rm tr}\big(k_1D_1-k_2 D_2\big),
\end{align}
where the vector superfield ${\cal V}$ is as defined in section \ref{sectionN=23} and
$\mu$ is the mass parameter.
In this action as well as in the component field
expansion of the original GT action \eqref{sfN=2GT}, the auxiliary fields $D_1$ and
$D_2$ appear linearly. Their equations of motion determine the
auxiliary scalar fields $\sigma_1$ and $\sigma_2$, respectively.
The effect of the above $D$-term deformation is then, shifting the values of $\sigma_1$ and
$\sigma_2$ as follows
\begin{align}\label{shift}
\sigma_1\to\sigma_1-\frac{\mu}{2},\quad\quad
\sigma_2\to \sigma_2+\frac{\mu}{2}.
\end{align}
After integrating out $D_1$ and $D_2$ using their equations of
motion, $\sigma_1$ and $\sigma_2$ appear only in the ${\cal L}_{\rm
ferm}^D$ and ${\cal L}_{\rm bos}^D$ terms of the GT Lagrangian. Therefore, the shifting
in \eqref{shift} affects only the $D$-term fermionic and bosonic potentials in
\eqref{N=2GT}. Explicitly,
we will obtain
 \begin{align}\label{D-term1}
{\cal L}_{\rm ferm}^D={\cal L}_{\rm ferm}^D({\rm GT})+\mu~{\rm
tr}\big(i\xi_A^\dagger\xi^A -i\omega^{\dagger A}\omega_A\big)
\end{align}
\begin{align}\label{D-term2}
{\cal L}_{\rm bos}^D=&{\cal L}_{\rm bos}^D({\rm GT})-{\rm
tr}\Big[\mu^2\big(Z_A^\dagger Z^A + W^{\dagger A}
W_A\big)-\frac{4\pi\mu}{k_1}\Big(Z^AZ_A^\dagger Z^BZ_B^\dagger -
W^{\dagger A} W_A W^{\dagger B} W_B\Big)
\nonumber\\
&~~~~~~~~~~~~~~~~~~~ -\frac{4\pi\mu}{k_2}\Big(Z_A^\dagger Z^A Z_B^\dagger Z^B - W_A
W^{\dagger A}W_B  W^{\dagger B}\Big)\Big].
\end{align}

In summary, the Lagrangian of the deformed theory in terms of the component fields
is written as
\begin{align}\label{massL}
{\cal L}_{\rm tot}={\cal L}_{\rm GT}+{\cal L}_{\rm ferm}^\mu+{\cal L}_{\rm bos}^\mu,
\end{align}
where the first term in the right hand side is the original GT
Lagrangian and the last two are the $D$-term  deformations in
\eqref{D-term1} and \eqref{D-term2}. It is important to note that
the $D$-term deformation does not affect the $F$-term potentials
where we have the crucial difference between the ${\cal N}=2$ and
${\cal N}=3$ theories. As a result, the supersymmetry preserving
mass deformation, which is derived from the $D$-term deformation,
has the same form for these two theories.

It is straightforward to show that the Lagrangian \eqref{massL} is invariant under the ${\cal N}=2$ supersymmetry
\eqref{N=2susy} if we include the following additional variations for the
fermionic fields,
\begin{align}
&\delta_\epsilon^\mu\xi^A = \mu\epsilon Z^A,\qquad
\delta_\epsilon^\mu\xi_A^\dagger = \mu\bar\epsilon Z_A^\dagger,
\nonumber \\
&\delta_\epsilon^\mu\omega_A = -\mu\epsilon W_A, \quad
\delta_\epsilon^\mu\omega^{\dagger A} = -\mu\bar\epsilon W^{\dagger A}.
\end{align}
Furthermore, in the special case of $c_i=\frac{2\pi}{k_i}$ the action \eqref{massL}
is invariant under the additional ${\cal N}=1$ supersymmetry
\eqref{susyN=3-1} and \eqref{susyN=3-2} with additional supersymmetry variations,
\begin{align}
&\delta_\eta^\mu\xi^A = -i\mu\eta W^{\dagger A},\quad
\delta_\eta^\mu\xi_A^\dagger = \mu\eta W_A,
\nonumber \\
&\delta_\eta^\mu\omega_A = -i\mu\eta Z^\dagger_A, \quad\,\,\,
\delta_\eta^\mu\omega^{\dagger A} = \mu\eta Z^A.
\end{align}

\subsection{$F$-term deformation}
An alternative realization of the mass deformation is in
terms of $F$-term superpotential deformation. In this section we will
show that such realization of the supersymmetry preserving mass
deformation of the GT theories is possible for ${\cal N}=2$ and ${\cal N}=3$ cases.
In the case of ${\cal N}=3$, the $F$-term deformation is equivalent to the $D$-term
deformation of the previous subsection up to field redefinition, while
for the ${\cal N}=2$ theory these deformations cannot be related by field redefinition.

In the ${\cal N}=2$ superfield language the $F$-term deformation is
given by
\begin{align}
S_F=-\mu\int d^3xd^2\theta~{\rm tr}\big({\cal Z}^A{\cal
W}_A\big) -\mu\int d^3xd^2\bar\theta~{\rm tr}\big(\bar {\cal Z}_A\bar {\cal W}^A\big).
\end{align}
Carrying out the component field expansion of the GT Lagrangian
including this $F$-term deformation,  we obtain
\begin{align}
{\cal L}_{\rm tot}={\cal L}_{\rm GT}+{\cal L}_F^\mu,
\end{align}
where
\begin{align}
{\cal L}_F^\mu=&\mu~{\rm tr}\big(\xi^A\omega_A-\omega^{\dagger
A}\xi^\dagger_A\big)
-\mu^2\big(Z^AZ^\dagger_A+W_AW^{\dagger A}\big)
\nonumber \\
&+2\mu~{\rm
tr}\Big[\big(c_1W_AZ^BW_B+c_2W_BZ^BW_A\big)W^{\dagger A}
+W_A\big(c_1W^{\dagger B}Z^\dagger_BW^{\dagger A}
+c_2W^{\dagger A}Z^\dagger_BW^{\dagger B}\big)
\nonumber\\
&~~~~~~~~~~+\big(c_1Z^BW_BZ^A+c_2Z^AW_BZ^B\big)Z^\dagger_A
+Z^A\big(c_1Z^\dagger_AW^{\dagger B}Z^\dagger_B+c_2Z^\dagger_BW^{\dagger B}
Z^\dagger_A\big)\Big].
\end{align}
In order to cast this term into the form of the
$D$-term deformation in \eqref{massL}, we introduce the following field
redefinitions
\begin{align}\label{fieldred}
&Z^A=\frac1{\sqrt 2}\big(P^A-Q^{\dagger A}\big),\quad\quad
W_A=\frac1{\sqrt 2}\big(P^{\dagger}_A+Q_A\big),
\nonumber\\
&\xi^A=\frac1{\sqrt 2}\big(\chi^A-i\eta^{\dagger A}\big),\quad\quad
\omega_A=\frac1{\sqrt 2}\big(i\chi^\dagger_A+\eta_A\big).
\end{align}
With this field redefinition we obtain
\begin{align}
{\cal L}_F^\mu=&\mu~{\rm tr}\big(i\chi_A^\dagger\chi^A
-i\eta^{\dagger A}\eta_A\big)-\mu^2{\rm tr}\big( P_A^\dagger P^A +
Q^{\dagger A} Q_A\big)
\nonumber\\
&-2\mu \,{\rm tr}\Big[c_1\big(P^AP_A^\dagger P^BP_B^\dagger - Q^{\dagger A} Q_A
Q^{\dagger B} Q_B\big) +c_2\big(P_A^\dagger P^A P_B^\dagger P^B
- Q_A Q^{\dagger A}Q_B  Q^{\dagger B} \big)\Big].
\end{align}
In the special case of $c_i=\frac{2\pi}{k_i}$, this Lagrangian is
equivalent to the $D$-term deformation \eqref{massL}.
In addition, one can show that the form of the original GT Lagrangian is
invariant under our field redefinition \eqref{fieldred}. Therefore, we realize that
in the case of ${\cal N}=3$ GT theory the $F$-term deformation is
equivalent to the supersymmetry preserving mass deformation derived from the $D$-term
deformation. However, in the case of ${\cal N}=2$, the $D$-term and $F$-term deformations
give two different supersymmetry preserving mass deformations.

\section{Vacua of the Mass-deformed GT Theories}

The classical supersymmetric discrete vacua of the the mass-deformed
ABJM theory were obtained in \cite{Gomis:2008vc} and refined in
\cite{Kim:2010mr}. In this section, we will follow a similar
procedure with \cite{Gomis:2008vc,Kim:2010mr} to obtain the
classical discrete vacua of the mass-deformed GT theory. The
structures of the vacua are the same as those of the mass-deformed
ABJM theory, except for overall coefficients which depend on the CS
levels $k_1$, $k_2$, and the size of the irreducible blocks inside
matrix representations of the vacua.

In the $D$-term deformed ${\cal N}=2,3$ GT theories the bosonic
potential can be written as\footnote{As pointed out earlier the
F-term deformation is equivalent to D-term deformation in ${\cal
N}=3$ theory. However, for ${\cal N}=2$ theory the two are not the
same and the vacuum equations for F-term deformed theory are
different from what we are considering here.}
\begin{align}\label{bospot}
V_{{\rm bos}}^\mu = |\sigma_1Z^A - Z^A\sigma_2 -\mu Z^A|^2 +
|\sigma_2 W_A - W_A\sigma_1 + \mu W_A|^2 + |F^A|^2 + |G_A|^2,
\end{align}
where $\sigma_1$ and $\sigma_2$ were defined in \eqref{auxilflds},
$F^A$ and $G_A$ are
\begin{align}
F^A= -2\big(c_1W^{\dagger B}Z_B^\dagger W^{\dagger A}
+c_2 W^{\dagger A}Z_B^\dagger W^{\dagger B}\big),
\quad
G_A= -2\big(c_1 Z_A^\dagger W^{\dagger B} Z_B^\dagger
+c_2 Z_B^\dagger W^{\dagger B} Z_A^\dagger\big).
\end{align}
Here we have introduced the notation $|{\cal O}|^2\equiv {\rm
tr}\,{\cal O}^\dagger {\cal O}$, for convenience. At the vacuum,
$V_{{\rm bos}}^\mu =0$. This means each of the summand in
\eqref{bospot} is vanishing separately.
Vanishing of the first two terms in the right hand side of \eqref{bospot} is rewritten as
\begin{align}\label{vacua-eqns}
&k_1 Z^AZ_B^\dagger Z^B + k_2 Z^B Z_B^\dagger Z^A = \frac{k_1k_2}{2\pi}\,\mu Z^A,
\nonumber \\
&k_1 W^{\dagger A} W_BW^{\dagger B} + k_2 W^{\dagger B}W_BW^{\dagger
A} =-\frac{k_1k_2}{2\pi}\,\mu W^{\dagger A}.
\end{align}
In order to solve the vacuum equation we assume that
 $\mu >0$, $k_1>0$, and $k_2<0$, without loss of generality.
Then the equations in \eqref{vacua-eqns} are simplified as
\begin{align}\label{res-vac-eqns}
&\tilde Z^A\tilde Z_B^\dagger \tilde Z^B
- t \tilde Z^B \tilde Z_B^\dagger \tilde Z^A + \tilde Z^A=0,
\nonumber \\
&\tilde W^{\dagger A} \tilde W_B\tilde W^{\dagger B}
- t \tilde W^{\dagger B}\tilde W_B\tilde W^{\dagger A}-\tilde W^{\dagger A}=0,
\end{align}
where $t= -\frac{k_2}{k_1}$  and we rescaled the fields as
\begin{align}
Z^A = \left(\frac{|k_2|\mu}{2\pi}\right)^{\frac12}\tilde Z^A, \quad
W^{\dagger A}=\left(\frac{|k_2|\mu}{2\pi}\right)^{\frac12}\tilde W^{\dagger A}.
\end{align}
\\

\noindent
\underline{{\bf ${\rm U}(1)_{k_1}\times{\rm U}(1)_{k_2}$} {\bf case}}

\noindent In the mass-deformed ABJM theory with $U(1)\times U(1)$
gauge group, there is only trivial vacuum solution $Z^A=W_A=0$.
However, the vacuum equation of the mass-deformed GT theory with
$U(1)\times U(1)$ gauge group is nontrivial and can be written as
\begin{align}\label{res-vac-eqns2}
|\tilde Z^1|^2+|\tilde Z^2|^2 =\frac1{t-1},\quad \tilde W_A=0, \quad
{\rm or}\quad|\tilde W_1|^2+|\tilde W_2|^2 =\frac1{1-t},\quad \tilde
Z^A=0.
\end{align}
Expanding the complex fields in terms of real fields as $\tilde Z^A
= \tilde X_A + i \tilde X_{A+4},\,\, \tilde W_A = \tilde X_{A+2}
-i\tilde X_{A+6}$, the vacuum equations in \eqref{res-vac-eqns2} are
given by
\begin{align}\label{res-vac-eqns3}
\tilde X_1^2+\tilde X_2^2 +\tilde X_5^2+\tilde
X_6^2=\frac1{t-1},\quad \tilde W_A=0, \quad {\rm or}\quad\tilde
X_3^2+\tilde X_4^2 +\tilde X_7^2+\tilde X_8^2=\frac1{1-t},\quad
\tilde Z^A=0.
\end{align}
For $t>1$  the first equation defines a $S^3$ while for $t<1$ the
second equation defines a $S^3$. We note that the vacuum equations
\eqref{res-vac-eqns2} are singular for $t=1$, as a result the $S^3$
vacuum modulus does not exist in the mass-deformed abelian ABJM
theory. \vskip 0.3cm
 \noindent \underline{{\bf ${\rm U}(N)_{k_1}\times{\rm
U}(N)_{k_2}$} {\bf case}}

\noindent For non-abelian GT theory, the solutions of
\eqref{res-vac-eqns} are represented by a direct sum of two sets of irreducible
rectangular matrices\footnote{We follow the notation introduced in
\cite{Kim:2010mr,Cheon:2011gv}.}. The first set is composed of
two $n\times (n+1)$ matrices,
\begin{align}\label{mat-1}
{\cal M}_1^{(n)}&=\frac{1}{\sqrt{(n+1) t -n}}\left(\begin{array}{cccccc}
\sqrt{n\!}\!\!\!&0&&&&\\&\!\sqrt{n\!-\!1} \!\!&\!0&&&\\
&&\ddots&\ddots&&\\&&&\sqrt{2}&0&\\&&&&1&0\end{array}\right),
\nonumber \\
{\cal M}_2^{(n)}&=\frac{1}{\sqrt{(n+1) t -n}}
\left(\begin{array}{cccccc}0&1&&&&\\&0&\sqrt{2}&&&\\ &&\ddots&\ddots&&\\
&&&0\!&\!\!\sqrt{n\!-\!1}\!&\\&&&&0&\!\!\!\sqrt{n\!}\end{array}
\right),
\end{align}
and the second one is composed of two $(n+1)\times n$ matrices,
\begin{align}\label{mat-2}
\bar{\cal M}_1^{(n)}&=\frac1{\sqrt{n+1-  nt}}
\left(\begin{array}{ccccc}\sqrt{n}\!\!\!&&&&\\0&\!\sqrt{n\!-\!1}&&&\\
&0&\ddots&&\\&&\ddots&\sqrt{2}&\\&&&0&1\\&&&&0\end{array}\right),
\nonumber \\
\bar{\cal M}_2^{(n)}&=\frac1{\sqrt{n+1-  nt}}
\left(\begin{array}{ccccc}0&&&&\\1&0&&&\\
&\sqrt{2}&\ddots&&\\&&\ddots&0&\\&&&\sqrt{n\!-\!1}\!&\!0\\&&&&\!\!\!\sqrt{n}
\end{array}\right).
\end{align}
${\cal M}_a^{(n)}$ and $\bar{\cal M}_a^{(n)}$ can be used to
construct the solutions of the first and the second equations in
\eqref{res-vac-eqns}, respectively. These irreducible blocks are
similar to those in \cite{Kim:2010mr,Cheon:2011gv}, except for the
overall coefficients.\footnote{These blocks are obtained following
the method of \cite{Gomis:2008vc}, where the uniqueness of the
irreducible blocks with $t=1$ was confirmed.   Therefore, the same
argument of uniqueness can be applied to the irreducible blocks in
\eqref{mat-1} and \eqref{mat-2}.} As a result of these non trivial
overall coefficients, there is an interesting difference between the
vacuum solution here and those of \cite{Kim:2010mr,Cheon:2011gv}.
For given $t<1$ the blocks ${\cal M}_{1,2}^{(n)}$ with
$n=\frac{t}{1-t}$ are not allowed while for $t>1$ the blocks
$\bar{\cal M}_{1,2}^{(n)}$ with $n=\frac1{t-1}$ are not allowed.
Therefore, for GT theories with CS levels satisfying any of these
two restrictions on the ratio $t$ of the CS levels, the total number
of classical supersymmetric vacua is reduced as compared to the ABJM
case.

The general solutions satisfying the equations \eqref{vacua-eqns} and the F-term equations $|F^A|=0$, $|G_A|=0$, are represented
in terms of the irreducible blocks as
\begin{align}\label{ZW-vacua}
Z^A&=\left(\frac{|k_2|\mu}{2\pi}\right)^{\frac12}\left(\begin{array}{c}
\begin{array}{cccccc}\mathcal{M}_A^{(n_1)}\!\!&&&&&\\&\!\!\ddots\!&&&&\\
&&\!\!\mathcal{M}_A^{(n_i)}&&& \\ &&& {\bf 0}_{(n_{i+1}+1)\times n_{i+1}}
&&\\&&&&\ddots&\\&&&&&{\bf 0}_{(n_f+1)\times n_f}\end{array}\\
\end{array}\right),\nonumber
\end{align}
\begin{align}
W^{\dagger A}&=\left(\frac{|k_2|\mu}{2\pi}\right)^{\frac12}\left(\begin{array}{c}
\begin{array}{cccccc}{\bf 0}_{n_1\times (n_1+1)}&&&&&\\&\ddots&&&&\\
&&{\bf 0}_{n_i\times(n_i + 1)} &&&\\
&&& \bar{\mathcal M}_A^{(n_{i+1})}\!\!&&\\&&&&\!\!\ddots\!&\\
&&&&&\!\!\bar{\mathcal M}_A^{(n_f)}\end{array}\\
\end{array}\right),
\end{align}
where $n_i =0,1,2,\cdots$ and ${\bf 0}_{m\times n}$ denotes $m\times n$ zero matrix.
Since $Z^A$ and $W^{\dagger A}$ are $N\times N$ matrices,
there are two constraints
\begin{align}
\sum_{n=0}^\infty\big[n \tilde N_n + (n+1) \hat N_n\big] = N,\quad
\sum_{n=0}^\infty\big[(n+1) \tilde N_n + n \hat N_n\big] = N,
\end{align}
where $\tilde N_n$ denotes the number of block of ${\mathcal
M}_A^{(n)}$-type and $\hat N_n$ is the number of block of
$\bar{\mathcal M}_A^{(n)}$-type. Here $\tilde N_0$ and $\hat N_0$
represent the numbers of empty columns and empty rows, respectively.

Next we discuss one more interesting feature of the classical
supersymmetric discrete vacua we found here. It is understood that
at the discrete vacua, the U(N)$\times$U(N) gauge symmetry is
partially broken. The unbroken gauge symmetry corresponds to the
reshuffling of irreducible blocks of the same type. More precisely,
the gauge fields for the symmetry that reshuffles the block ${\cal
M}_A^{(n)}$ is given by~\cite{Kim:2010mr}
\begin{align}\label{amu}
A_\mu &= {\bf 1}_n \otimes[a_\mu]_{\tilde N_n\times\tilde N_n},
\nonumber \\
\hat A_\mu &= {\bf 1}_{n+1}\otimes [a_\mu]_{\tilde N_n\times\tilde N_n},
\end{align}
where $n=0,1,2,\cdots$ and $a_\mu$ denotes the unbroken gauge field
generating the reshuffling of $\tilde N_n$ blocks. Inserting
\eqref{amu} into the CS action \eqref{LLCS} we obtain
\begin{align}
&\frac{1}{4\pi}\epsilon^{\mu\nu\rho}\left(k_1\,{\rm tr}{\bf 1}_n
+ k_2\,{\rm tr}{\bf 1}_{n+1}\right)
{\rm tr}_{\tilde N_n}\left[a_\mu\partial_\nu a_\rho + \frac{2i}{3}a_\mu a_\nu a_\rho\right]
\nn \\
&=\frac{k_1 n + k_2(n+1)}{4\pi}\epsilon^{\mu\nu\rho}
{\rm tr}_{\tilde N_n}\left[a_\mu\partial_\nu a_\rho + \frac{2i}{3}a_\mu a_\nu a_\rho\right].
\end{align}
Similarly for the blocks $\bar {\cal M}_A^{(n)}$ the unbroken gauge
fields are
\begin{align}
A_\mu &= {\bf 1}_{n+1} \otimes[a_\mu]_{\hat N_n\times\hat N_n},
\nonumber \\
\hat A_\mu &= {\bf 1}_{n}\otimes [a_\mu]_{\hat N_n\times\hat N_n},
\end{align}
and the corresponding CS term is given by
\begin{align}
\frac{k_1 (n+1) + k_2 n}{4\pi}\epsilon^{\mu\nu\rho}
{\rm tr}_{\hat N_n}\left[a_\mu\partial_\nu a_\rho + \frac{2i}{3}a_\mu a_\nu a_\rho\right].
\end{align}
In summary, for a given vacuum there exist CS theories with gauge
group
\begin{align}
\prod_{n=0}^{\infty}{\rm U}(\tilde N_n)_{\tilde k_n}\times {\rm U}(\hat N_n)_{\hat k_n},
\end{align}
where  CS levels are
\begin{align}
\tilde k_n =k_1 n + k_2(n+1)=nF_0 + k_2,\qquad
\hat k_n = k_1 (n+1) + k_2 n=nF_0 + k_1.
\end{align}
Here we would like to point out two interesting facts about the
unbroken gauge group. The first point is that unlike the
mass-deformed ABJM theory, where the CS levels of the unbroken gauge
fields are unshifted i.e., $\tilde k_n = -k$ and $\hat k_n =
k$~\cite{Kim:2010mr}, in the present case the CS levels are shifted
by $nF_0$. The other point is that when $n=\frac{t}{1-t}$
 or $n=\frac1{t-1}$ the CS levels, $\tilde
k_n$ or $\hat k_n$ are vanishing, respectively. This is consistent
with what we explained previously about the coefficients of the
irreducible blocks.\footnote{We are indebted to Seok Kim for
clarifying this point.}

\section{Conclusion}

In this paper we found the maximal supersymmetry preserving mass
deformation of the ${\cal N}=2,3$ GT theories. Since the original GT
Lagrangian was written in the ${\cal N}=2$ superfield formalism, the
additional ${\cal N}=1$ supersymmetry is not manifest in the case of
the ${\cal N}=3$ GT theory. To clarify this point we started with
the component field expansion of the GT Lagrangians. Then we wrote
the ${\cal N}=2$ supersymmetry transformation rules for the
component fields. Gaiotto and Tomasiello pointed out that when the
coefficients of the superpotential $c_i=\frac{2\pi}{k_i}$, the
supersymmetry is enhanced to ${\cal N}=3$~\cite{Gaiotto:2009mv}. We
found the explicit supersymmetry transformation rules for the
component fields under the additional ${\cal N}=1$ supersymmetry.

Following the line of the original ABJM theory we found the mass
deformations of the GT theories which preserve the maximal
supersymmetry. The mass deformations can be realized either as
D-term or F-term deformations. The $D$-term deformation does not
affect the $F$-term potentials. Since the ${\cal N}=2,3$ theories
differ by the $F$-term potential, the mass deformation which is
derived from the $D$-term deformation for the two theories are
equivalent. On the other hand, for the ${\cal N}=3$ GT theory the
$F$-term deformation is equivalent to the mass deformation obtained
from the $D$-term deformation, up to field redefinition, while the
${\cal N}=2$ $D$-term and $F$-term deformations give two distinct
supersymmetry preserving mass deformations.

Using the mass-deformed GT theories we found set of discrete
classical supersymmetric vacua. The classical vacuum solutions are
expressed in terms irreducible blocks of size $n\times(n+1)$ or
$(n+1)\times n$. An important feature of the current situation is
that when the ratio of the CS levels $t=\frac {n+1}n$ or $t=\frac
n{n+1}$, those irreducible blocks have singular coefficients and are
not allowed. Therefore, in mass-deformed GT theories with the ratio
of CS levels satisfying the above conditions the total number of
supersymmetric vacua is reduced.

The purpose of this work is mainly to clear the way for future
perspective in this subject. There are many unanswered questions in
the GT theory. The important ones are the facts that M-theory
interpretation and M-theory limit of dual gravity are not
understood. Recently, Cheon {\it et al} \cite{Cheon:2011gv} found  a
one-to-one correspondence between the supersymmetric vacua in gauge
theory side and $\mathbb{Z}_k$ quotients of the LLM geometry. This
correspondence is valid for weakly-curved as well as strongly-curved
background in dual gravity. Based on the close relation between the
discrete vacua of the mass-deformed ABJM and GT theories, there is a
possibility that the investigation of \cite{Cheon:2011gv} (see also
\cite{Hashimoto:2011nn}) can be carried out for mass-deformed GT
theories as well. The result can shed some light on the physics of
strongly-curved and strongly-coupled massive type IIA string theory
in relation with the conjecture of \cite{Aharony:2010af}.

\section*{Acknowledgements}
The authors would like to thank Min-Young Choi and Seok Kim for
useful discussions. This work was supported by the Korea Research
Foundation Grant funded by the Korean Government  with grant number
2009-0073775, 2011-0009972 (O.K.), and 2009-0077423 (D.D.T.).

\appendix

\section{Conventions and Fierz Identity }\label{CV-UF}
We choose (2+1)-dimensional gamma matrices as $ \gamma^0= i\sigma^2,
\gamma^1=\sigma^1$, and $\gamma^2=\sigma^3$, which satisfy
$\gamma^\mu\gamma^\nu=\eta^{\mu\nu} +
\epsilon^{\mu\nu\rho}\gamma_\rho$. The conventions for spinor
indices are\footnote{We use the spinor convention of
\cite{Benna:2008zy}. There is one difference in the convention of
the suppressed spinor indices, i.e., in our case $\xi\gamma^\mu\chi
=\xi^\alpha\gamma_{\alpha}^{\mu\,\beta}\chi_\beta$, where $\xi$ and
$\chi$ are two component spinors.}
\begin{align}
&\theta^\alpha = \epsilon^{\alpha\beta}\theta_\beta,\quad
\theta_\alpha = \epsilon_{\alpha\beta}\theta^\beta,\quad
\epsilon^{12}=-\epsilon_{12}=1,
\nonumber \\
&\theta^\alpha\theta_\alpha\equiv\theta^2,\quad
\theta^\alpha\bar\theta_\alpha = \theta\bar\theta, \quad
\theta^\alpha\gamma_\alpha^{\mu\beta}\bar\theta_\beta\equiv
\theta\gamma^\mu\bar\theta.
\end{align}
In terms of these conventions for spinors, we obtain
\begin{align}
\theta_\alpha\theta_\beta=
\frac12\epsilon_{\alpha\beta}\theta^2,\quad
\theta^\alpha\theta^\beta = -\frac12\epsilon^{\alpha\beta}\theta^2.
\end{align}
Useful Fierz identities inside the trace  are
\begin{align}
{\rm tr}\big[(\epsilon\gamma^\mu\psi_A)(\psi_B\gamma_\mu\psi_C)\big]
&= -{\rm tr}\big[2(\epsilon\psi_C)(\psi_A\psi_B) +
(\epsilon\psi_A)(\psi_B\psi_C)\big]
\nonumber \\
&=-{\rm tr}\big[(\epsilon\psi_C)(\psi_A\psi_B) -
(\epsilon\psi_B)(\psi_C\psi_A)\big]
\nonumber \\
&={\rm tr}\big[(\epsilon\psi_A)(\psi_B\psi_C) +2
(\epsilon\psi_B)(\psi_C\psi_A)\big],
\nonumber \\
{\rm tr}\big[(\psi_A\gamma^\mu\psi_B)(\psi_C\gamma_\mu\epsilon)\big]
&= -{\rm tr}\big[2(\psi_B\psi_C)(\psi_A\epsilon) +
(\psi_A\psi_B)(\psi_C\epsilon)\big]
\nonumber\\
&=-{\rm tr}\big[(\psi_B\psi_C)(\psi_A\epsilon) -
(\psi_C\psi_A)(\psi_B\epsilon)\big]
\nonumber \\
&={\rm tr}\big[(\psi_A\psi_B)(\psi_C\epsilon) +2
(\psi_C\psi_A)(\psi_B\epsilon)\big],
\nonumber \\
{\rm tr}\big[(\epsilon\psi_A)(\psi_B\psi_C)\big] &= -{\rm tr}\big[
(\epsilon\psi_B)(\psi_C\psi_A) +
(\epsilon\psi_C)(\psi_A\psi_B)\big],
\end{align}
where $\epsilon$ is a spinor without gauge indices. We also have the
relations,
\begin{align}
(\theta\bar\theta)^2 = -\frac12\theta^2\bar\theta^2,\quad
\theta\bar\theta (\theta\gamma^\mu\bar\theta)=0,\quad
(\theta\gamma^\mu\bar\theta)(\theta\gamma^\nu\bar\theta)=\frac12\eta^{\mu\nu}
\theta^2\bar\theta^2.
\end{align}
We adapt the convention for integrations,
\begin{align}
&d^2\theta\equiv -\frac14d\theta^\alpha d\theta_\alpha,\qquad
d^2\bar\theta\equiv-\frac14 d\bar\theta^\alpha
d\bar\theta_\alpha,\quad d^4\theta\equiv d^2\theta d^2\bar\theta,
\nonumber \\
& \int d^2\theta\theta^2 =1,\quad \int d^2\bar\theta\bar\theta^2
=1,\quad \int d^4\theta\theta^2\bar\theta^2=1,
\end{align}
and have supercovariant derivatives and supersymmetry generators,
\begin{align}
&D_\alpha = \partial_\alpha - i
\gamma^{\mu\beta}_\alpha\bar\theta_\beta\partial_\mu,\quad \bar
D_\alpha = -\bar\partial_\alpha -
i\theta^\beta\gamma_{\beta\alpha}^\mu\partial_\mu,
\nonumber \\
&Q_\alpha =\partial_\alpha + i
\gamma^{\mu\beta}_\alpha\bar\theta_\beta\partial_\mu,\quad \bar
D_\alpha = -\bar\partial_\alpha +
i\theta^\beta\gamma_{\beta\alpha}^\mu\partial_\mu
\end{align}
with anti-commutation relations,
\begin{align}
&\{D_\alpha,\,\bar
D_\beta\}=-2i\gamma_{\alpha\beta}^\mu\partial_\mu,\qquad
\{Q_\alpha,\,\bar Q_\beta\}= 2i\gamma_{\alpha\beta}^\mu\partial_\mu.
\end{align}

\end{document}